\documentclass[10pt,twocolumn,a4paper,final]{IEEEtran}
\usepackage{indentfirst}
\usepackage{setspace}
\usepackage[numbers,sort&compress]{natbib}
\usepackage{graphicx}
\usepackage{amstext}
\usepackage{algorithm}
\usepackage{algorithmic}
\usepackage{mathrsfs}
\usepackage{amssymb}
\usepackage{amsmath}
\usepackage{epstopdf}
\usepackage{CJK}
\usepackage{multicol}
\usepackage{stfloats}
\usepackage{url}
\usepackage{color}
\usepackage{enumerate}
\usepackage{diagbox}
\usepackage{tabularx}
\usepackage{footnote}
\makesavenoteenv{tabular}
\allowdisplaybreaks[4]
\ifCLASSINFOpdf
\else
\fi
\begin{document}
%
\title{\huge{Large Intelligent Surface (LIS)-based Communications: \\New Features and System Layouts}}
%
%
%



\author{
\IEEEauthorblockN{Jide Yuan\IEEEauthorrefmark{1}, Hien Quoc Ngo\IEEEauthorrefmark{1}, and Michail Matthaiou\IEEEauthorrefmark{1}}

\IEEEauthorblockA{\IEEEauthorrefmark{1}Institute of Electronics, Communications and Information Technology (ECIT), Queen's University Belfast, U.K. e-mail: $\rm \{y.jide, hien.ngo, m.matthaiou\}@qub.ac.uk$}
}

\maketitle

\pagestyle{empty}  
\thispagestyle{empty} 

\begin{spacing}{1}
\begin{abstract}

The concept of large intelligent surface (LIS)-based communication has recently attracted increasing research attention, where a LIS is considered as an antenna array whose entire surface area is available for radio signal transmission and reception.
In order to provide a fundamental understanding of \emph{LIS-based} communication, this paper studies the uplink performance of LIS-based communication with matched filtering in the presence of a line-of-sight channel.
We first study the new features introduced by LIS. In particular, the array gain, spatial resolution, and the capability of interference suppression are theoretically presented and characterized.
Then, we study two possible LIS system layouts, i.e., centralized LIS (C-LIS) and distributed LIS (D-LIS), and propose a user association scheme aiming to maximize the minimum user spectral efficiency (SE).
Simulation results compare the achievable SE between two system layouts. We observe that the proposed user association algorithm significantly improves the performance of D-LIS, and with the help of it, the per-user achievable SE in D-LIS outperforms that in C-LIS in most considered scenarios.

\end{abstract}
\end{spacing}


%
\IEEEpeerreviewmaketitle
\newtheorem{Definition}{Definition}
\newtheorem{Lemma}{Lemma}
\newtheorem{Theorem}{Theorem}
\newtheorem{Corollary}{Corollary}
\newtheorem{Proposition}{Proposition}
\newtheorem{Remark}{Remark}
\newtheorem{Property}{Property}
\newcommand{\rl}[1]{\color{red}#1}

%
%
%
%

%

\section{Introduction}
\begin{spacing}{1}

\emph{Large intelligent surface} (LIS) is a future man-made structure envisioned in \cite{abs190406704}, which integrates an uncountably number of new-form antennas (or modules) into a limited surface area, forming a spatially continuous surface. A LIS has great ability to manipulate electromagnetic waves, and can theoretically make the entire wireless communication environment intelligent \cite{cui2017}.

The \emph{LIS-based} communication concept can be regarded as an extension of traditional massive multi-input multi-output (mMIMO), in which the LIS architecture uses its entire surface for transmission and reception.
This concept was firstly proposed in \cite{8108330}, in which the uplink (UL) rate was evaluated for a near-field scenario. The result showcases a novel feature, namely that, the multiplexing capability of \emph{LIS-based} system is essentially determined by the wavelength $\lambda$.
Further, as extensions of \cite{8108330}, the authors studied the potential of LIS for positioning and derived the Cram\'{e}r-Rao lower bound for positioning a terminal on the central perpendicular line was derived in closed-form \cite{8264743}.
The authors in \cite{abs181005667} investigated the feasibility of splitting a large LIS into a number of small LIS-units. The asymptotic UL rate is given for multiuser scenarios, which shows that a \emph{LIS-based} system can achieve a comparable performance with conventional mMIMO with a significantly reduced area for antenna deployment.
However, the existing literature focuses on analyzing the near-field propagation environment, while research on \emph{LIS-based} systems for far-field scenarios, a more common scenario, is currently missing.

To give a full picture of \emph{LIS-based} communication concept, this paper studies the system from an UL perspective with match-filtering (MF) in the far-field.
Firstly, several new properties of LIS architecture are introduced. The results reveal the fact that array gain, and spatial resolution of a LIS architecture are highly dependent on the LIS size and frequency band.
With the new features introduced by LIS, we investigate the achievable spectral efficiency (SE) of \emph{LIS-based} communication in multiuser scenario with two different layouts, i.e., centralized LIS (C-LIS) and distributed LIS (D-LIS).
To fully reap the potential of D-LIS, we propose a large scale fading (LSF)-based user association scheme for maximizing the minimum user SE.
The simulation results reveal that the proposed user association scheme can significantly boost the system performance in terms of the minimum user SE. More importantly, with the help of the proposed user association scheme, D-LIS topologies offer a much higher coverage probability compared with C-LIS ones.

\section{System Model}

Consider a two-dimensional circular LIS deployed on the $xy$-plane and $K$ single-antenna users located in a three-dimensional space. For ease of understanding and in order to investigate the fundamental properties, we first consider a typical scenario, in which the LIS center is located at $x=y=z=0$, while the users are located at the space $z>0$.
We assume a LoS-dominated far-field propagation scenario where the path loss between a particular user to every point on the LIS is the same.\footnote{This assumption is reasonable for the following two reasons. Firstly, LISs are naturally deployed much higher above the sea level, making the signal strength from LoS path significantly stronger than that from scattering paths. More importantly, as clarified in \emph{Remark} 1 later in this paper, with the MF applied at the LIS architecture, the effective channel from a LoS path and scattering path are nearly orthogonal due to the capability of interference suppression, which results in a LoS-dominated propagation environment.} Specifically, we consider the distance between the $k$th user located at $\left(x_k,y_k,z_k\right)$ to the LIS center as the effective distance, which is given by
\begin{equation}\label{d_c_k}
{d_{k}^c} = \sqrt {{x_k^2} + {y_k^2} + {z_k^2}}.
\end{equation}

The free-space path loss is then expressed as a function of distance between transmit and receive antennas, and is given by \cite{molisch2004ieee}
\begin{equation}\label{PL_k}
{\text{P}}{{\text{L}}_{k}} = \left( {{{1}} \mathord{\left/
 {\vphantom {{{1}} {2\kappa {d_k^c}}}} \right.
 \kern-\nulldelimiterspace} {2\kappa {d_k^c}}}\right)^2,
\end{equation}
respectively, where $\kappa  = \frac{2\pi }{{\lambda }}$ with $\lambda$ being the wavelength. Note that the far-field path loss is valid when $d^c_k$ is larger than \emph{Fraunhofer distance}, i.e., $d^c_k>\frac{8R^2}{\lambda}$, where $R$ is the radius of the circular LIS.




\vspace{-0.2cm}

\subsection{Channel Model}


For a far-field scenario, since the distance between a user and LIS is sufficient large, we assume that the angles-of-arrival (AoA) for a user to each point at the LIS are identical. Therefore, the general channel propagation from the $k$th user to the point $(x,y,0)$ at a typical LIS can be represented as
\begin{equation}\label{g_los_k}
{g_{k}}\left( {x,y} \right) = {\text{PL}}_k^{\frac{1}{2}} \cdot {h_{k}}\left( {x,y} \right),
\end{equation}
where
\begin{equation}\label{h_los_k}
{h_{k}}\left( {x,y} \right) = {e^{ - j\left( {\kappa {d_k} + {\varphi _k}} \right)}},
\end{equation}
and $\varphi_k$ is random phase following uniform distribution in the range of $[-\pi,\pi]$,
${d_k}$ is the distance between the $k$th user and the point $\left(x,y,0\right)$ at LIS, and is given by
\begin{equation}\label{d_k}
{d_{k}} = \sqrt {{z_k^2} + {{\left( {x - {x_k}} \right)}^2} + {{\left( {y - {y_k}} \right)}^2}}.
\end{equation}
Note that minor differences between the distance from a particular user to any two points at the LIS will hardly impact path loss but will substantially impact the phase of $h_k(x,y)$. Therefore, we treat the effect of distance on the path loss and phase shift separately.

%
%

\vspace{-0.2cm}

\subsection{Effective Channel and Achievable SE with MF Scheme}

Based on (\ref{h_los_k}), the received signal at the LIS location $(x, y, 0)$ from all $K$ terminals is given by
\begin{equation}\label{}
r\left( {x,y} \right) = \sum\nolimits_{k = 1}^K {\sqrt {{p_k}} {g_{k}}\left( {x,y} \right){s_k}}  + n\left( {x,y} \right),
\end{equation}
where $s_k$ is the transmitted signal of the $k$th user with $\left\|s_k\right\|^2 = 1$, and $n\left( {x,y} \right)$ is the additive white Gaussian noise at LIS with variance ${\sigma ^2}$.
Applying the MF at the LIS, the received signal for the $k$th user is given as
\begin{align}\label{}
{r_k}\left( {x,y} \right)= \sum\nolimits_{k' = 1}^K {\sqrt {{\text{P}}{{\text{L}}_k}{\text{P}}{{\text{L}}_{k'}} {p_{k'}}}} {\Sigma^{\mathcal{S}} _{kk'}}{s_k} + {\omega _{k}},
\end{align}
where the effective channel
\begin{align}\label{Coeffi_kk}
{\Sigma^{\mathcal{S}} _{kk'}} \triangleq \iint_{\left( {x,y} \right) \in {\mathcal{S}}} {h_{k}^ * \left( {x,y} \right){h_{k'}}\left( {x,y} \right)dxdy},
\end{align}
where ${\mathcal{S}}$ is the surface-area of LIS; ${\omega _k}$ is the noise after MF with zero-mean and variance
\begin{equation}\label{noise_p}
{\mathsf{E}}\left[ {\omega _k^ * {\omega _{k}}} \right] = {\text{P}}{{\text{L}}_k}{\Sigma^{\mathcal{S}}_{kk}}{\sigma ^2}.
\end{equation}

With the effective channel given in (\ref{Coeffi_kk}) and the noise model in (\ref{noise_p}), the achievable SE for the $k$th user is then calculated as
\begin{equation}\label{Rate_general}
{\mathsf{R}_k}  = {\log _2}\left( {1 + \frac{{{p_k}{\text{P}}{{\text{L}}_k}{{\left( {\Sigma _{kk}^\mathcal{S}} \right)}^2}}}{{\Sigma _{kk}^\mathcal{S}{\sigma ^2} + \sum\nolimits_{k' \ne k} {{p_{k'}}{\text{P}}{{\text{L}}_{k'}}{{\left| {\Sigma _{kk'}^\mathcal{S}} \right|}^2}} }}} \right).
\end{equation}

\section{Intrinsic Properties of LIS Architectures}

In this section, we investigate the effective channel to illustrate the new features introduced by our LIS architecture.

\begin{figure}[!t]
	\centering
	\includegraphics[width=6cm]{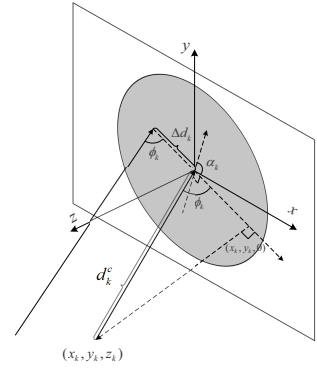}
	\caption{The radiating model of a transmitting signal to a circular LIS in far-field scenarios where the channel phase shift can be evaluated using $\triangle d_k$. }
\end{figure}

To evaluate $\Sigma^{\mathcal{S}}_{kk'}$ with circular LIS, it is necessary to measure the wave phase at each point on the LIS. We first project the radiation direction of the $k$th user on the $xy$-plane, as shown in Fig. 1. Since the AoAs $\phi_k$ for the $k$th user to each point at LIS are nearly identical, and it is clear that for any lines that are perpendicular to the projection line, the wave phases of points on the line are also identical. Hence, denoting by $\alpha_k$ the angle of elevation (AoE), and consider the wave phase on the zero crossings line
\begin{equation}\label{line_k}
{\ell _k} = \left\{ {\left( {x,y} \right):y - x\tan {\alpha _k} = 0} \right\}
\end{equation}
as reference, the phase at point $(x,y,0)$ can be calculated via its distance to ${\ell _k}$, and the channel response is then rewritten as
\begin{equation}\label{}
{h_{k}}\left( {x,y} \right) ={e^{ - j\left( {\kappa d_k^c + \Delta d_k\kappa \cos {\phi _k} + {\varphi _k}} \right)}},
\end{equation}
where
\begin{equation}\label{delta_k}
\Delta d_k \triangleq \frac{{y - x\tan {\alpha _k}}}{{\sqrt {{{\tan }^2}{\alpha _k} + 1} }}.
\end{equation}

\begin{Definition}
We now define ${\eta _{kk'}}$ and ${\xi _{kk'}}$, which have the following structure
\begin{equation}\label{eta_kk}
{\eta _{kk'}} = {{{x_k}} \mathord{\left/
 {\vphantom {{{x_k}} {d_k^c}}} \right.
 \kern-\nulldelimiterspace} {d_k^c}} - {{{x_{k'}}} \mathord{\left/
 {\vphantom {{{x_{k'}}} {d_{k'}^c}}} \right.
 \kern-\nulldelimiterspace} {d_{k'}^c}},
\end{equation}
\begin{equation}\label{xi_kk}
{\xi _{kk'}} = {{{y_k}} \mathord{\left/
 {\vphantom {{{y_k}} {d_k^c}}} \right.
 \kern-\nulldelimiterspace} {d_k^c}} - {{{y_{k'}}} \mathord{\left/
 {\vphantom {{{y_{k'}}} {d_{k'}^c}}} \right.
 \kern-\nulldelimiterspace} {d_{k'}^c}},
\end{equation}
and
\begin{equation}\label{chi_kk}
{\chi^2 _{kk'}} = \eta _{kk'}^2 + \xi _{kk'}^2.
\end{equation}
\end{Definition}

As shown in (\ref{angle2cor1}) and (\ref{angle2cor2}) in Appendix A and with the fact that $\sin\phi_k={z_k}/{d^{c}_k}$, all the coefficients, i.e., ${\eta _{kk'}}$, ${\xi _{kk'}}$, and ${\chi _{kk'}}$, can be expressed as a function of AoA and AoE. The use of coordinates is for its simplicity and intuitive.


\begin{Proposition}
The effective channel $\Sigma^{\mathcal{S}}_{kk'}$ of a circular LIS equals
\begin{equation}\label{}
{\Sigma^{\mathcal{S}}_{kk'}} = {{\rm A}_{kk'}} \cdot {\rm B} \left(R,\kappa,{\chi _{kk'}}\right),
\end{equation}
where
\begin{equation}\label{}
{{\rm A}_{kk'}} = {e^{j\left( {\kappa \left( {d_k^c - d_{k'}^c} \right) + {\varphi _k} - {\varphi _{k'}}} \right)}},
\end{equation}
\begin{equation}\label{}
{\rm B} \left(R,\kappa,{\chi _{kk'}}\right)= 2\pi R\frac{{{J_1}\left( {R\kappa {\chi _{kk'}}} \right)}}{{\kappa {\chi _{kk'}}}},
\end{equation}
where $J_n(\cdot)$ is the $n$th order of Bessel function of the first kind.
\end{Proposition}
\begin{IEEEproof}
See Appendix A.
\end{IEEEproof}

Note that ${{\rm A}_{kk'}}$ is a constant phase shift which depends on the users' positions and original phase, while ${\rm B} \left(R,\kappa,{\chi _{kk'}}\right)$ is the \emph{LIS response} with MF which reveals the interference suppression and the spatial resolution of the LIS.

\subsection{Array Gain}

Array gain relates to the received signal power at the LIS \cite{8319526}. 
We have the following property for the array gain.

\begin{Property}
$\Sigma^{\mathcal{S}}_{kk'}$ represents the array gain when $k'=k$, which equals to
\begin{equation}\label{}
\Sigma^{\mathcal{S}}_{kk} = \pi R^2.
\end{equation}
\end{Property}
\begin{IEEEproof}
It is intuitive that ${\rm{A}}_{kk}=1$, and by leveraging \emph{L'Hospital's Rule}, we have
\begin{equation}\label{}
\mathop {\lim }\limits_{x \to 0} \!\frac{{{J_1}\!\left(\! {ux}\! \right)}}{x}\!=\!\mathop {\lim }\limits_{x \to 0}\! \frac{{\partial {J_1}\!\left( \!{ux} \!\right)}}{{\partial x}} \!\mathop  = \limits^{(a)}\! {\left. {\!\frac{u}{2}\!\left(\! {{J_0}\left( {ux} \right) \!-\! {J_2}\left( {ux} \right)} \right)} \right|_{x = 0}},
\end{equation}
where $(a)$ is obtained from \cite[Eq. 03.01.20.0006.01]{123456789}. Noting that $J_0(0)=1$ and $J_2(0)=0$, we complete the proof.
\end{IEEEproof}

The result holds for any circular LIS with finite surface-area.
The conclusion that the array gain equals to the surface area makes intuitive sense, which is, to some extent, similar to the conventional antenna array whose gain in a LoS environment approaches the number of elements in the antenna array \cite{1424539}.
Fig. 2a shows the absolute value of the effective channel ${\Sigma}_{kk'}^{\mathcal{S}}$ with respect to $\chi_{kk'}$, in which the $\chi_{kk'}=0$ represents the array gain of LIS.
A more remarkable insight is that with a larger LIS, ${\Sigma}_{kk'}^{\mathcal{S}}$ converges more quickly with respect to $\chi_{kk'}$, which indicates that the effective channels from users further apart are almost orthogonal.

\subsection{Spatial Resolution}

The spatial resolution represents the minimum related distance of two users so that the ratio of the interference to the array gain is smaller than a predefined and small threshold. A precise definition of the spatial resolution is as follows.
\begin{Definition}
By denoting $\tilde\Sigma^{\mathcal{S}}_{kk'}=\left|\frac{\Sigma^{\mathcal{S}}_{kk'}}{\Sigma^{\mathcal{S}}_{kk}}\right|$,
the spatial resolution $\bar\chi$ is then defined as:
\emph{For any two users $k$ and $k'$ whose $\chi_{kk'}>\bar\chi$, $\tilde\Sigma^{\mathcal{S}}_{kk'} <\eta$, where $\eta$ is a small positive value.}
\end{Definition}

%


\begin{Property}
The function $f(ux)=\frac{{{J_1}\left( {ux} \right)}}{{ux}}$ converges to 0 with increasing $x$.
Specifically, the function exhibits damped oscillation, and each of the local minima and maxima is a constant value which is uncorrelated to scale parameter $u$, and is given by $\frac{{{J_1}\left( {{j_{2,n}}} \right)}}{{{j_{2,n}}}}$, for $n \in {\mathbb{N}_ + }$,
where $j_{m,n}$ is the $n$th zero of the $J_m(x)$ function. Moreover, the absolute value of $\frac{{{J_1}\left( {{j_{2,n}}} \right)}}{{{j_{2,n}}}}$ decreases with respect to $n$.
\end{Property}
\begin{IEEEproof}
The convergence of the function is obvious by recalling the well-known feature that the radius of convergence of the Bessel function of the first kind is infinite. Then,
by differentiating the function with respect to $x$ in range $x>0$, and let the result equal to 0, we have
\begin{equation}\label{}
\frac{\partial }{{\partial x}}\left( {\frac{{{J_1}\left( {ux} \right)}}{{ux}}} \right) \mathop  = \limits^{(b)}  - \frac{{{J_2}\left( {ux} \right)}}{x}=0,
\end{equation}
where $(b)$ is due to \cite[Eq. 03.01.20.0009.01]{123456789}. The roots of the equation are simply obtained as $
{x_n} = \frac{{{j_{2,n}}}}{u}$ for $n \in {\mathbb{N}_ + }$.
Substituting $x_n$ into $f(ux)$, we obtain the formula of each maxima and minima. Along with the fact that the absolute value of local minima and maxima of $J_1(\cdot)$ monotonically decreases with $n$, we complete the proof.
\end{IEEEproof}


With the property given above, we can observe that $\left|f(ux)\right|$ cannot reach $\frac{2{\left| {{J_1}\left( {{j_{2,n}}} \right)} \right|}}{{{j_{2,n}}}}$ again when $x> {{j_{2,n}}}$. This phenomenon perfectly matches the definition of $\bar\chi$ since when we choose $\frac{{\left| {{J_1}\left( {{j_{2,n}}} \right)} \right|}}{{{j_{2,n}}}}$ as $\eta$ and ${j_{2,n}}$ as $\bar\chi$, for any $x>j_{2,n}$, we have ${\rm \tilde B} \left({R,\kappa,\chi _{kk'}}\right)<\eta$. Therefore, we obtain the spatial resolution criterion in the following proposition.

\begin{Proposition}
Set $\eta_n = \frac{2{\left| {{J_1}\left( {{j_{2,n}}} \right)} \right|}}{{{j_{2,n}}}}$ as our threshold, the spatial resolution of a circular LIS with respect to the its radius is given as ${\chi _{kk'}} = \frac{{{j_{2,n}}}}{{\kappa R}}$, for $n \in {\mathbb{N}_ + }$, where $n$ is adjustable according to the resolution requirement.
\end{Proposition}

From the above expression, we clearly observe that the spatial resolution increases with $\kappa$ and $R$. Fig. 2b shows $\tilde\Sigma^{\mathcal{S}}_{kk'}$ with respect to $R$, which verifies our analysis. It can be seen that the absolute value of the normalized response monotonically decreases in an oscillatory manner. Moreover, with larger size, the LIS is able to avail of higher spatial resolution.
Similar conclusion has been drawn in \cite{8319526}, where with sufficient large surface area of the LIS, any two users can be almost separated without interference, even if they are located very close together.


\begin{figure}[!t]
	\centering
	\includegraphics[width=9cm]{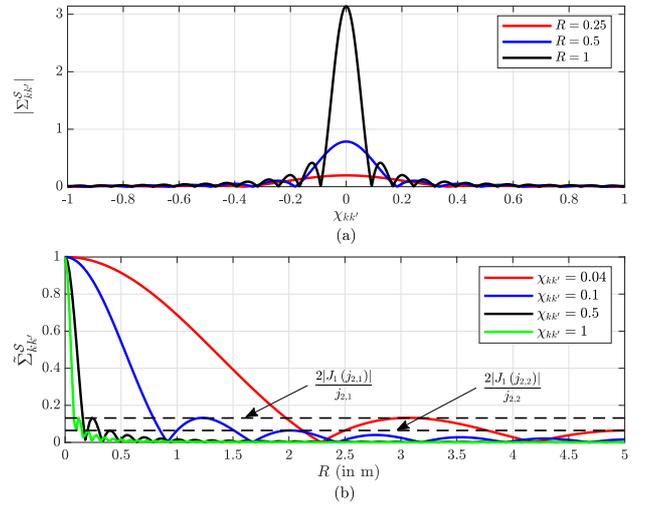}
	\caption{$\left|\Sigma^{\mathcal{S}}_{kk'}\right|$ and $\tilde\Sigma^{\mathcal{S}}_{kk'}$ with respect to $\chi_{kk'}$ and $R$, respectively.}
\end{figure}

\begin{Remark}
Assume that a scatterer is placed besides a user, and further assume that the signal from the LoS path and the reflected signal off the scatterer reach the LIS simultaneously. In this case, the reflected signal can be regarded as a LoS signal transmitted from an image of the user created by the scatterer with weaker strength.
\end{Remark}


%

\section{Performance Analysis of C-LIS and D-LIS}

In this section, we study the achievable SE of a LIS system in different topological layouts, i.e., C-LIS and D-LIS. 


\subsection{C-LIS system}

We consider a C-LIS system, in which the LIS can intelligently change the surface area that is being used for transmission.
The system performance of such an architecture is then derived in the following proposition.

\begin{Proposition}
In a C-LIS, the achievable SE of the $k$th user is described as
\begin{equation}\label{R_k}
{\mathsf{R}_k} = {\log _2}\left( {1 + \frac{{{p_k}{\text{P}}{{\text{L}}_k}}}{{\frac{1}{{\pi {R^2}}}{\sigma ^2} + \sum\nolimits_{k' \ne k} {{p_{k'}}{\text{P}}{{\text{L}}_{k'}}{\rm{\tilde B}}{{\left( {R,\kappa ,{\chi _{kk'}}} \right)}^2}} }}} \right),
\end{equation}
where ${\rm \tilde B} \left(R,\kappa,{\chi_{kk'}}\right)= \frac{{\rm B} \left(R,\kappa,{\chi _{kk'}}\right)}{{\rm B} \left(R,\kappa,{\chi_{kk}}\right)}$,
and the overall sum SE across $K$ users then equals
\begin{equation}\label{R_k_sum}
{\mathsf{R}^{\text{C-LIS}}_{{\text{total}}}} = \sum\nolimits_{k = 1}^K {{\mathsf{R}_{\text{k}}}}.
\end{equation}
\end{Proposition}
\begin{IEEEproof}
The result can be directly obtained based on \emph{Proposition} 1 and (\ref{Rate_general}).
\end{IEEEproof}

The result in (\ref{R_k}) reveals the joint impact of the LIS radius, wavelength, and users' positions on the achievable SE, which indicates that LIS is able to maximize ${\mathsf{R}_k}$ by adjusting the frequency band and its size.

\begin{Corollary}
In a C-LIS, if $R$ is sufficient large or/and the frequency is high, the achievable SE of the $k$th user is approximated as
\begin{equation}\label{R_k_inf}
{\mathsf{R}_k}  \approx {\log _2}\left( {1 + \frac{{p_k}}{{{\sigma ^2}}}{\pi {R^2}}{\text{P}}{{\text{L}}_k}} \right),
\end{equation}
and the overall sum SE across $K$ users then equals
\begin{equation}\label{R_k_sum_inf}
{\mathsf{R}^{\text{C-LIS}}_{{\text{total}}}} \approx \sum\nolimits_{k = 1}^K {{{\log }_2}\left( {1 + \frac{{p_k}}{{{\sigma ^2}}}{\pi {R^2}}{\text{P}}{{\text{L}}_k}} \right)} .
\end{equation}
\end{Corollary}
\begin{IEEEproof}
When the size of LIS is sufficient large or the frequency band is sufficient high, according to \emph{Property} 2, it can infer that that the channel response ${\rm{B}}\left( {R,\kappa ,{\chi _{kk'}}} \right)$ normalized by array gain $\pi R^2$ satisfies the relationship
\begin{equation}\label{Interf_inf}
\mathop {\lim }\limits_{R \to \infty } {\rm{\tilde B}}\left( {R,\kappa ,{\chi _{kk'}}} \right) = \mathop {\lim }\limits_{\kappa  \to \infty } {\rm{\tilde B}}\left( {R,\kappa ,{\chi _{kk'}}} \right) = 0.
\end{equation}
Substituting (\ref{Interf_inf}) into (\ref{R_k}), we complete the proof.
\end{IEEEproof}

\begin{figure}[!t]
	\centering
	\includegraphics[width=9.5cm]{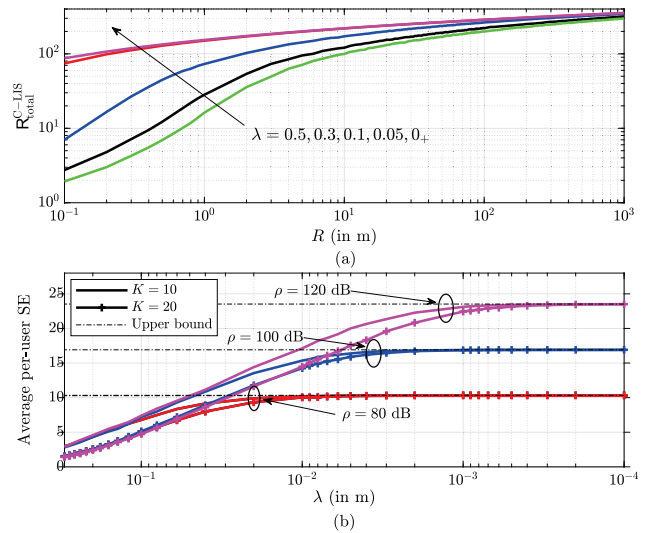}
	\caption{Performance of C-LIS, in which (a) compares the $\mathsf{R}^{\text{C-LIS}}_{\text{total}}$ against $R$ with $\rho = 100$\,dB, and (b) compares the per-user SE with respect to $\lambda$. The results are shown for $\sigma^2 = -174$\,dBm/Hz/s, and averaged over 100 runs.}
\end{figure}


The result shows a similar conclusion as in conventional mMIMO, in which when the number of antennas at BS is infinite, the interference can be fully canceled.
Moreover, we observe an advantage of LIS compared with conventional mMIMO, which is the great capability of interference suppression at high frequency bands. In other words, the result in (\ref{R_k_sum_inf}) can be regarded as an upper bound of the achievable SE.
Fig. 3 verifies our theoretical analysis, where the curve $\lambda = 0_+$ represents the upper bound in (\ref{R_k_sum_inf}). We assume the same transmit power across $10$ users, and denote $\rho = \left\{\frac{p_k}{\sigma^2}\right\}_{k=1,\ldots,K}$.\footnote{The parameter $\rho$ refers to the ratio of the signal power at the transmitter to the noise power at LIS. The reason that using $\rho$ instead of the conventional SNR is that, with randomly deployed users, the SNR at the LIS is unknown due to the path loss.}
From Fig. 3a, it can be seen that the achievable sum SE is closer to the upper bound with a higher frequency band or with a larger LIS surface.
An important observation is that, even when $R$ increases to 1000\,m, for the case that $\lambda = 0.3$\,m (i.e., 1\,GHz), there still remains a small gap to the upper bound in (\ref{R_k_sum_inf}), which indicates that interference should not be ignored in microwave bands. Moreover, as there is hardly any gap between $\lambda=0.05$\,m and the upper bound, the expressions in (\ref{R_k_inf}) and (\ref{R_k_sum_inf}) can perfectly approximate the performance for mmWave band, even with finite LIS.
Fig. 3b compares the per-user achievable SE in regard to $\lambda$ for $K=10$ and $K= 20$ scenarios.
We can observe that the per-user achievable SEs can reach the upper bounds in any scenarios when the wavelength is sufficient short.
More importantly, the per-user SE for $10$\,users and $20$\,users converge to the same value, which indicates the great advantage of LIS architecture for meeting the massive connectivity requirements of 6G networks.

\subsection{D-LIS system}


To deploy a centralized LIS spanning tens or even hundreds of square meters is not always feasible; for this reason, we hereafter consider a distributed topology, in which $M (M>K)$ same-size LIS-units are spread over a large area and are connected to a centralized baseband unit.
In D-LIS systems, each LIS-unit serves a particular user.
In this section, we study the achievable SE for a D-LIS, and we aim to design effective schemes to maximize the minimum user SE by proposing new user association, orientation control and power control schemes.


\subsubsection{SE Analysis}


We first evaluate the SE of the $k$th user achieved at arbitrary LIS-unit, e.g., the $m$th LIS-unit.

\begin{Proposition}
Assuming the $m$th LIS-unit is assigned to the $k$th user, the achievable SE is then given as
\begin{equation}\label{Dis_per_rate}
{\mathsf R}_k^{m}\!=\!{\log _2}\!\left(\! {1 + \frac{{{p_k}{\text{P}}{{\text{L}}_{k,m}}}}{{\frac{1}{{\pi R_{\rm{d}}^2}}{\sigma ^2} + \sum\limits_{k' \ne k} {{p_{k'}}{\text{P}}{{\text{L}}_{k',m}}{\rm{\tilde B}}{{\left( {{R_{\rm{d}}},\kappa ,\chi _{m,kk'}} \right)}^2}} }}} \!\right),
\end{equation}
where $R_{\rm{d}}$ is the radius of the LIS-unit in a D-LIS topology, ${\text{P}}{{\text{L}}_{k,m}}$ is the path loss from the $m$th LIS-unit to the $k$th user, and $\chi _{m,kk'}$ has
the same form as in \emph{Proposition} 3 by simply using the coordinates of the $m$th LIS-unit.
\end{Proposition}
\begin{IEEEproof}
The result can be directly obtained by substituting \emph{Definition} 1 and \emph{Proposition} 3 into (\ref{Rate_general}).
\end{IEEEproof}

Different from the expression in (\ref{R_k_sum}) in which the achievable SEs of each user couple together due to ${\rm{\tilde B}}\left( {R,\kappa ,\chi _{kk'} } \right)$, the per-user achievable SE at each LIS-unit is independent across $M$ LIS-units, which allow us to adjust each unit separately.

\subsubsection{User Association}

By harnessing the result in \emph{Proposition} 4, we elaborate on user association for minimum user SE maximization. Denote by $\bf S$ the $K\times M$ LIS selection matrix, whose $(k,m)$th element is $s_{k,m}\subset[0,1]$ with $s_{k,m} =1$ representing that the $k$th user is associated to the $m$th LIS, and $s_{k,m} =0$ otherwise.
Given the achievable SE expression in (\ref{Dis_per_rate}), we formulate the minimum SE maximization problem as
\begin{subequations}\label{SR_Ori}
\begin{align}
\label{ob_ori_mmax}\mathop {{\text{maximize}}}\limits_{\mathbf{S}}\;\;& {\mathop {\min }\limits_k \left\{ {{s_{k,m}}\mathsf{R}_k^m} \right\}}\\
\label{select_ele}{\rm{s}}{\rm{.t}}{\rm{.}}\;\;&{s_{k,m}} \subset \left[ {0,1} \right],\;\;\;\;\;\;\forall k,\;m,\\
\label{select_row}&\left\| {\bf s}_k \right\|_0  = 1,\;\;\;\;\;\;\;\;\;\forall k,\\
\label{select_column}&\left\| \left\{{\bf S}\right\}_m \right\|_0 \subset [0,1], \;\;\forall m,
\end{align}
\end{subequations}
where ${\bf s}_k$ and $\left\{{\bf S}\right\}_m$ represent the $k$th row and the $m$th column of $\bf S$, respectively. The constraint (\ref{select_row}) ensures that each user is served by a LIS, and (\ref{select_column}) guarantees the each LIS serves no more than one user.
Note that, the minimum user SE maximization problem is nonconvex even without discrete constraints, whose optimal result can only be solved via searching. 
Therefore, we propose a suboptimal iterative user association algorithm to reduce this complexity.

By noting the fact that interference can be largely reduced by adjusting the orientation of the LIS, i.e., ${\rm{\tilde B}}{{\left( {R_{\rm{d}},\kappa ,{\chi_{m,kk'}}} \right)}}$ can be ignored when $k \neq k'$, a large scale fading (LSF)-based user association (LUA) scheme is then proposed. When no interference is considered, the maximization problem in (\ref{SR_Ori}) can be rewritten in the following form
\begin{subequations}\label{MMax_Ori}
\begin{align}\label{R_mmax_var}\mathop {{\text{maximize}}}\limits_{\mathbf{S}} \;\;&{\mathop {\min }\limits_k \left\{ {{s_{k,m}}{\text{P}}{{\text{L}}_{k,m}}} \right\}} \\
\label{select_same}{\rm{s}}{\rm{.t}}{\rm{.}}\;\;\;&(\ref{select_ele}),(\ref{select_row}),(\ref{select_column}),
\end{align}
\end{subequations}
respectively. Even when adopting the LUA, the problems are still not solvable since the constraints in (\ref{select_same}) are discrete, which makes the optimization problems be non-convex. Hence, we design a reweighted $\ell_1$-norm iterative method to approximate the constraint \cite{Cands2008}. As the $s_{k,m}$ is either 0 or 1, we hence approximate the coefficient in the following form
\begin{equation}\label{app_s}
\left\|{s}_{k,m}\right\|_0 \approx \left\|\omega_{k,m}{\tilde s}_{k,m}\right\|_1,
\end{equation}
where ${\tilde s}_{k,m} \in [0,1]$ is a continuous value, and $\omega_{k,m}=\frac{1}{{\tilde s}_{k,m}+\varrho}$ denotes the weight coefficient associated with ${\tilde s}_{k,m}$, in which $\varrho$ is a very small positive value that provides stability. It is straightforward to see that the right hand of (\ref{app_s}) will force the expression converge to either $0$ or $1$. Utilizing this approximation, the problem (\ref{R_mmax_var}) can finally be transformed as
\begin{subequations}\label{SR_var_var}
\begin{align}
\label{ob_var_mmax}\mathop {{\text{maximize}}}\limits_{\mathbf{S}}\;\;&{\mathop {\min }\limits_k \left\{ {{{\omega}_{k,m}{\tilde s}_{k,m}}{\text{P}}{{\text{L}}_{k,m}}} \right\}}\\
\label{select_ele_var}{\rm{s}}{\rm{.t}}{\rm{.}}\;\;&{{\tilde s}_{k,m}} \in \left[ {0,1} \right],\;\;\;\;\;\;\;\;\;\;\forall k,\;m,\\
\label{select_row_var}&\left\|{\boldsymbol\omega}_{k}\cdot{\tilde {\bf s}}_{k} \right\|_1  = 1,\;\;\;\;\;\;\;\;\;\;\;\forall k,\\
\label{select_column_var}&\left\| \left\{{\bf{\Omega}\cdot \bf S}\right\}_m \right\|_1 \in [0,1], \;\;\forall m,
\end{align}
\end{subequations}
where ${\bf{\Omega}} \in \mathbb{R}^{K\times M}$ is the weight matrix whose $(k,m)$th element is $\omega_{k,m}$, and ${\boldsymbol\omega}_{k}$ represents the $k$th column of $\bf{\Omega}$. Note that, with this approximation, the optimization problem (\ref{SR_var_var}) is convex with fixed $\bf \Omega$, we hence can develop an iterative method to achieve suboptimal LIS selection, where $\omega_{k,m}$ in each iteration is updated via the solution of ${\tilde s}_{k,m}$ from the previous iteration. According to the complexity analysis in \cite{gass2003linear}, the arithmetic complexity per iteration of our algorithm is $\mathcal{O}(K^{3.5})$. If we set the limit of the iterations as $N$, the overall complexity is then upper bounded by $\mathcal{O}(NK^{3.5})$. The procedure of LUA is detailed in \textbf{Algorithm} 1.

\begin{figure}[!b]
	\centering
	\includegraphics[width=9cm]{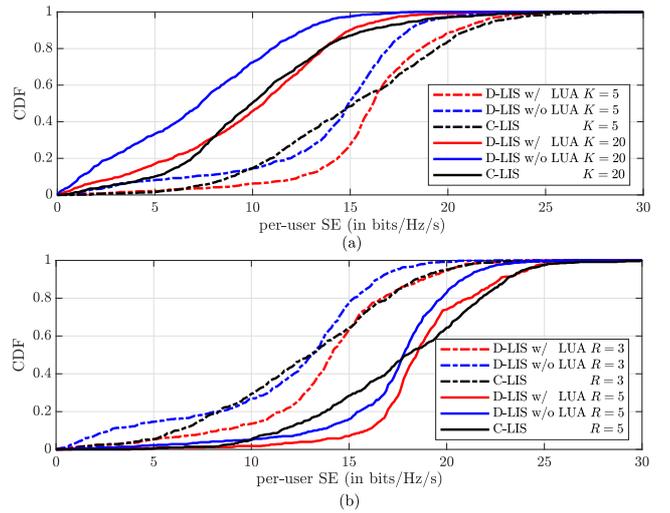}
	\caption{The CDFs of achievable SE are evaluated for C-LIS and D-LIS, in which (a) compares the CDFs under different users with $R=5$, and (b) compares the CDFs under different surface areas with $K=5$. The results are shown for $M=20$.}
\end{figure}

\newcommand{\algorithmicinitial}{\textbf{Initialization:}}
\newcommand{\algorithmicstep}{\textbf{Step}}
\newcommand{\STEP}{\item[\algorithmicstep]}
\newcommand{\INITIAL}{\item[\algorithmicinitial]}

\begin{algorithm}[!t]\label{algorithm1}
\caption{}
\begin{algorithmic}
\INITIAL Weight matrix ${\bf \Omega} = {\bf 1}^{K\times M}$, iteration count $Count = 1$, maximum iteration number $N$, and the parameter $\varrho$

\WHILE{$Count<=N$}
\STATE Solve problem (\ref{SR_var_var}).
\STATE Update $\omega_{k,m}$ via $\omega_{k,m}=\frac{1}{{\tilde s}_{k,m}+\varrho}$,
\STATE $Count=Count +1$
\ENDWHILE
\end{algorithmic}
\end{algorithm}


To illustrate the performance of C-LIS and D-LIS, we compare the CDFs of achievable per-user SE for different scenarios in Fig. 4. We assume that, in C-LIS, a large LIS is located at the center of a square of size 1\,$\text{km}^2$, while 20\,LIS-units are randomly deployed in the same size of area in D-LIS, and the overall surface area equals to the area of LIS in C-LIS for the sake of fairness.
The LUA aiming to maximize the minimum SE is also depicted in Fig. 4.
From Fig. 4a, an obvious observation is that the proposed LUA algorithm is more effective in the scenarios with fewer users. For example, when $K=5$, the 95\%-likely achievable SE for D-LIS with LUA is about 10\,bits/Hz/s which is 8\,bits/Hz/s higher than for D-LIS without LUA, while such advantage reduces to 1\,bits/Hz/s for the 20\,user case. More importantly, we observe that C-LIS outperforms D-LIS for the 20\,user scenario while the opposite situation occurs with $K=5$.
The reason is that, with more users, a higher spatial resolution LIS is required to distinguish the users which are closely located.
Fig. 4b shows the CDF of the achievable SE with different surface area for $K=5$.
We can see that by increasing the radius of surface from $3$\,m to $5$\,m, the 95\%-likely per-user SE can be improved over 200\%.
Besides, we clearly observe that, by applying LUA, D-LIS is superior to C-LIS for both $R=3$\,m and $R=5$\,m scenarios in terms of the per-user achievable SE, and such advantage is more significant with larger surface area.

\section{Conclusions}

We have considered a \emph{LIS-based} communication system, in which LIS is viewed as an antenna array that can be used for transmission and reception. With MF used at the LIS, we have shown that the array gain and the spatial resolution of LIS architecture is proportional to its surface area and radius, respectively.
We have investigated the UL performance of C-LIS and D-LIS, and designed effective LSF-based user association scheme for D-LIS for maximizing the minimum SE.
The simulation results reveal that the proposed algorithm can effectively enhance the system performance of D-LIS, and more importantly, we observe that D-LIS outperforms C-LIS in most of scenarios in terms of minimum SE with the help of a proposed association algorithm. This result indicates that a D-LIS deployment, which forms a wireless communication platform by multiple LIS-units with small surface area, is a more promising topology.

\appendices

\section{Proof for Proposition 1}

According to the analysis from (\ref{line_k}) to (\ref{delta_k}), the coefficient in (\ref{Coeffi_kk}) can be calculated as
\begin{align}\label{app_1}
&{\Sigma^{\mathcal{S}}_{kk'}}= \iint_{\left( {x,y} \right) \in \mathcal{S}} {h_{k}^*\left( {x,y} \right){h_{k'}}\left( {x,y} \right)dxdy} \hfill \notag\\
&= {{\rm A}_{kk'}} \iint_{\left( {x,y} \right) \in \mathcal{S}} {{e^{j\kappa \left( {\Delta {d_k}\cos {\phi _k} - \Delta {d_{k'}}\cos {\phi _{k'}}} \right)}}dxdy}\notag\\
& = {{\rm A}_{kk'}} \cdot {\rm B} \left({R,\kappa,\chi _{kk'}}\right).
\end{align}
Since ${{\rm A}_{kk'}}$ is only dependent on the user's position, we thus focus on deriving ${\rm B} \left({R,\kappa,\chi _{kk'}}\right)$.
Substituting (\ref{delta_k}) into (\ref{app_1}), we get
\begin{align}\label{B_kk}
&{\rm B} \left({R,\kappa,\chi _{kk'}}\right) \notag \\
&= \iint_{\left( {x,y} \right) \in \mathcal{S}}{\cos \left( {\kappa \Delta {d_k}\cos {\phi _k} - \kappa \Delta {d_{k'}}\cos {\phi _{k'}}} \right)dxdy} \hfill \notag \\
&= \iint_{\left( {x,y} \right) \in \mathcal{S}}{\cos \left( {\left( {\frac{{\kappa \cos {\phi _k}}}{{\sqrt {{{\tan }^2}{\alpha _k} + 1} }} - \frac{{\kappa \cos {\phi _{k'}}}}{{\sqrt {{{\tan }^2}{\alpha _{k'}} + 1} }}} \right)y} \right.}\notag\\
& \left. { - \left( {\frac{{\kappa \cos {\phi _k}\tan {\alpha _k}}}{{\sqrt {{{\tan }^2}{\alpha _k} + 1} }} - \frac{{\kappa \cos {\phi _{k'}}\tan {\alpha _{k'}}}}{{\sqrt {{{\tan }^2}{\alpha _{k'}} + 1} }}} \right)x} \right)dxdy.
\end{align}
Recalling the definition of $\tan{\alpha_k}$ and
$\cos{\phi_k}= \frac{\sqrt{x_k^2+y_k^2}}{d_k^c}$, it is easy to observe that
\begin{align}\label{angle2cor1}
\frac{{\cos {\phi _k}}}{{\sqrt {{{\tan }^2}{\alpha _k} + 1} }} = \frac{{{y_k}}}{{d_k^c}},
\end{align}
and
\begin{align}\label{angle2cor2}
\frac{{\cos {\phi _{k}}\tan {\alpha _{k}}}}{{\sqrt {{{\tan }^2}{\alpha _{k}} + 1} }} = \frac{{ {x_k}}}{{d_k^c}}.
\end{align}
Substituting \emph{Definition} 1 into (\ref{B_kk}), we can simplify the above integration as
\begin{align}\label{B_kk_1}
{\rm B} \left({R,\kappa,\chi _{kk'}}\right) = \iint\limits_{\left( {x,y} \right) \in \mathcal{S}} {\cos \left( {\kappa \left( {{\eta _{kk'}}y - {\xi _{kk'}}x} \right)} \right)dxdy}.
\end{align}
Further, we transform the integration (\ref{B_kk_1}) into polar coordinates, and the original integration is equivalent to
\begin{align}\label{B_kk_integ}
{\rm B}\!\left({R,\kappa,\chi _{kk'}}\right)&\!= \! \int\limits_0^R \!{\int\limits_0^{2\pi }\! r{\cos \left( { r\kappa\left( {{\eta _{kk'}}\sin \vartheta  - {\xi _{kk'}}\cos \vartheta } \right)} \right)d\vartheta dr} }  \notag\\
&\mathop  = \limits^{(a)} 2\pi \int_0^R {r{J_0}\left( {r\kappa{{\chi _{kk'}}} } \right)dr}\notag\\
& \mathop  = \limits^{(b)}2\pi R\frac{{{J_1}\left( {R\kappa {\chi _{kk'}}} \right)}}{{\kappa {\chi _{kk'}}}},
\end{align}
where $(a)$ is obtained by substituting (\ref{xi_kk}) into the integral, while $(b)$ can be obtained via \cite[Eq (6.521.1)]{gradshteyn2007ryzhik}.

\end{spacing}

\ifCLASSOPTIONcaptionsoff
  \newpage
\fi



%
%
%

\footnotesize
\bibliographystyle{IEEEtran}
\bibliography{myreference}

%

%
%
%




\end{document}